\documentclass[aps,prb,twocolumn,showpacs]{revtex4}
\usepackage{graphicx}
\usepackage{color}
\usepackage{bm}

\begin{document}

\title{Effect of an inhomogeneous exchange field on the proximity effect in disordered superconductor-ferromagnet hybrid structures}
\author{T. Champel and M. Eschrig}
\affiliation{
Institut f\"{u}r Theoretische Festk\"{o}rperphysik,
  Universit\"{a}t Karlsruhe,
 76128 Karlsruhe, Germany }
\date{\today}

\begin{abstract}
We investigate the effect of an inhomogeneous exchange field on 
the proximity effect in superconductor-ferromagnet hybrid structures within the quasiclassical theory of superconductivity.
As an example we study a superconductor-ferromagnet bilayer with an in-plane spiral magnetic order in the ferromagnet. 
The superconducting proximity effect induces in this case
triplet correlations in the bilayer
that are sensitive to the local quantization axis of the exchange field in the ferromagnet.
The coexistence of singlet and triplet pair correlations in the bilayer results into a sensitivity of
the superconducting transition temperature on the spatial variation of the exchange field in the ferromagnetic layer.
We show that the inhomogeneity also tends to suppress the oscillating behavior of the pair amplitudes in the ferromagnet. 
\end{abstract}


\maketitle

\section{Introduction}

The coexistence of superconductivity and magnetism is a long-standing issue and has gained recently a lot of attention due to 
experimental progress. One example is the discovery of ferromagnetic superconductors \cite{Sax2000,Hux2001}. 
Typically, magnetism tends to suppress singlet superconductivity via
two mechanisms of pair-breaking : 
(i) the orbital pair-breaking effect by action on the electron charges; 
(ii) the paramagnetic pair-breaking effect by action on the electron spin via the Zeeman coupling.

In contrast to the above-mentioned example of ferromagnetic superconductors,
in hybrid superconductor/ferromagnet (S/F) structures there is 
in general no coexistence of the magnetic and 
superconducting long range orders 
(assuming the ferromagnetic exchange field vanishes in S and the pairing interaction is repulsive or vanishingly small in F).
Nevertheless,
the influence of the magnetism on the superconductivity manifests itself near the S/F interface through the superconducting
proximity effect: Cooper pairs can penetrate a certain distance into the
ferromagnetic material.
As a result of the exchange splitting of the Fermi surface in the ferromagnet,
the Cooper pairs acquire a finite momentum and this causes spatial 
oscillations of the pair wave function in the F part. 
The oscillation of the pair amplitude  in F shares many similarities \cite{Dem1997} in its origin with the Fulde-Ferrell-Larkin-Ovchinnikov oscillations \cite{Ful1964,Lar1965} of the superconducting order parameter predicted to occur in the systems where the paramagnetic pair-breaking mechanism is dominant.
In addition, the pair-breaking effect of the exchange field results in a much shorter penetration of singlet superconducting pair correlations in the ferromagnet than in a nonmagnetic metal. 

In S/F bilayer systems a non-monotonic behavior is met again in the dependence of the superconducting critical temperature 
\cite{Rad1991,Fom2002} and in the dependence of the Josephson critical current \cite{Buz1982,Buz1991,Blu2002} on the ferromagnet thickness $d_{f}$ or on the amplitude of the exchange field (for a recent review see Ref.~\onlinecite{Lyu2004} and references therein).
Most of the quantitative investigations of these non-monotonic behaviors are based on the 
quasiclassical theory of superconductivity,\cite{Eil1968,Usa1970}
which provides the simplest framework to study the inhomogeneity of pair correlations near the S/F interface.
In the vaste majority of the theoretical works the exchange field  ${\bf J}$ in the F layer is considered homogeneous for the sake of simplicity.
However in reality 
often the magnetic system is characterized by the presence of 
an inhomogeneous magnetization, leading to a domain structure.
In this paper we address specifically the influence of such domain walls
on the proximity effect.

Recently, Rusanov {\em et al.}  \cite{Rus2004} have observed  in S/F bilayers
a quantitative dependence of the superconducting critical temperature $T_{c}$ on the domain state of the ferromagnet. In the presence of domain walls, $T_{c}$ is found to be 
enhanced compared to the $T_{c}$ obtained in the absence of domain walls. 
As the domain walls in Ref. \onlinecite{Rus2004} are argued to be of the
N\'{e}el type with an in plane magnetic moment, the
orbital pair-breaking effect is here negligible.
The  importance of the domain walls for the proximity effect has also been pointed out in Ref. \onlinecite{Lan2003}.

The above-mentioned effect is reminiscent to a similar effect in
magnetic superconductors with an inhomogeneous exchange field
(for a review see Ref. \onlinecite{Bul1985}, and references therein).
As was first hypothesized by Matthias and Suhl \cite{Mat1960}, 
one expects that 
the Cooper pairs experience an exchange field
averaged over the superconducting coherence length ($\xi_{s}$) scale, 
which leads effectively to a reduced pair-breaking effect 
for domain wall sizes comparable to or smaller than $\xi_{s}$.
It is however worth mentioning that this qualitative picture is borrowed 
from the physics of coexistence of magnetism and superconductivity, 
and its applicability stricto sensu to hybrid structures may be questioned. 
One goal of this paper is to point out and investigate in depth the mechanism responsible for the sensitivity of the (singlet) superconductivity on the directional changes of the exchange field in the S/F hybrid structures. 

The consideration of a full domain structure, i.e. with an alternation of regions with fixed magnetizations and domain walls represents a formidable theoretical task. 
This calls naturally for the study of simplified models in a preliminary step.
One possibility is to study a local inhomogeneity of ${\bf J}$ in the vicinity of the S/F interface. It has been found \cite{Ber2001,Kad2001,Ber2001b} that superconducting triplet correlations with an unusually long penetration length
in the ferromagnet arise through the S/F proximity effect. 
Such long range triplet correlations have been shown also to be
characteristic for F/S/F trilayer structures with non-collinear F moments.\cite{Vol2003,Ber2003}
It was concluded \cite{Ber2001,Kad2001,Ber2001b,Vol2003,Ber2003} from these two models that the sufficient ingredient for the existence of the long-range triplet components is a change in direction of the F moment.

\begin{figure}[t,b]
\begin{center}
\begin{picture}(0,0)%
\includegraphics{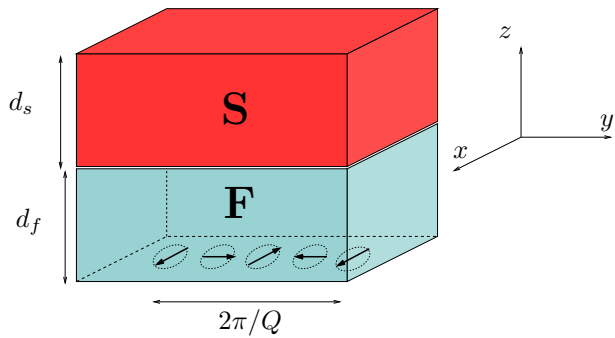}%
\end{picture}%
\setlength{\unitlength}{1243sp}%
\begingroup\makeatletter\ifx\SetFigFont\undefined
\def\x#1#2#3#4#5#6#7\relax{\def\x{#1#2#3#4#5#6}}%
\expandafter\x\fmtname xxxxxx\relax \def\y{splain}%
\ifx\x\y   
\gdef\SetFigFont#1#2#3{%
  \ifnum #1<17\tiny\else \ifnum #1<20\small\else
  \ifnum #1<24\normalsize\else \ifnum #1<29\large\else
  \ifnum #1<34\Large\else \ifnum #1<41\LARGE\else
     \huge\fi\fi\fi\fi\fi\fi
  \csname #3\endcsname}%
\else
\gdef\SetFigFont#1#2#3{\begingroup
  \count@#1\relax \ifnum 25<\count@\count@25\fi
  \def\x{\endgroup\@setsize\SetFigFont{#2pt}}%
  \expandafter\x
    \csname \romannumeral\the\count@ pt\expandafter\endcsname
    \csname @\romannumeral\the\count@ pt\endcsname
  \csname #3\endcsname}%
\fi
\fi\endgroup
\begin{picture}(12027,6447)(451,-6946)
\put(586,-4876){\makebox(0,0)[lb]{\smash{\SetFigFont{10}{12.0}{rm}{\color[rgb]{0,0,0}$d_{f}$}%
}}}
\put(451,-2536){\makebox(0,0)[lb]{\smash{\SetFigFont{10}{12.0}{rm}{\color[rgb]{0,0,0}$d_{s}$}%
}}}
\put(4726,-4696){\makebox(0,0)[lb]{\smash{\SetFigFont{17}{20.4}{rm}{\color[rgb]{0,0,0}\textbf{F}}%
}}}
\put(4681,-2806){\makebox(0,0)[lb]{\smash{\SetFigFont{17}{20.4}{rm}{\color[rgb]{0,0,0}\textbf{S}}%
}}}
\put(4636,-6901){\makebox(0,0)[lb]{\smash{\SetFigFont{10}{12.0}{rm}{\color[rgb]{0,0,0}$2 \pi /Q$}%
}}}
\put(10216,-1096){\makebox(0,0)[lb]{\smash{\SetFigFont{10}{12.0}{rm}{\color[rgb]{0,0,0}$z$}%
}}}
\put(9316,-3481){\makebox(0,0)[lb]{\smash{\SetFigFont{10}{12.0}{rm}{\color[rgb]{0,0,0}$x$}%
}}}
\put(12241,-2851){\makebox(0,0)[lb]{\smash{\SetFigFont{10}{12.0}{rm}{\color[rgb]{0,0,0}$y$}%
}}}
\end{picture}
\vspace{-0.2cm}
\caption{The exchange field rotates in the ferromagnetic layer in the plane $(xy)$ with a constant wavevector $Q$. The period of rotation in the $y$ direction  is $2\pi/Q$.} \label{Fig1}
\end{center}
\end{figure}

In a recent paper,\cite{Cha2004} we have studied an S/F bilayer within a 
model of an in-plane rotating magnetization 
(spiral order \cite{note}) in the F layer (see Fig.~\ref{Fig1}). 
This model is consistent with the fact that  the domain structure is expected a priori to appear in the layer rather than across the layer. 
We found \cite{Cha2004} that the long-range triplet 
components are not induced, although the exchange field is inhomogeneous.
To understand this peculiarity, it is worth noting that 
the models of moment inhomogeneities studied in the Ref. \onlinecite{Ber2001,Kad2001,Ber2001b,Vol2003,Ber2003} led
 to one-dimensional spatial dependences of the pair correlations: the inhomogeneity of the moment was always considered across the layers, i.e. competing with the inhomogeneity due to the proximity effect itself. 
Contrary to this,
the moment inhomogeneity within the in-plane spiral order model
occurs in the transverse direction, so that
the problem is intrinsically two-dimensional.

As pointed out in Ref.~\onlinecite{Ber2000}, 
the anomalous Green function $f$ and the Green function $g$, which are $2 
\times  2$ matrices in spin space, acquire both a nontrivial spin structure through the proximity effect. 
It has been realized \cite{Ber2003,Ber2004,You2004,Cha2004} that this always results in short-range triplet components 
(i.e. even in absence of inhomogeneity of ${\bf J}$) induced together with the singlet component in the ferromagnet near the S/F interface. 
For example, the difference in the $T_{c}$ in
F/S/F trilayers in the parallel and antiparallel configurations of the F moments \cite{Tag1999} stems precisely from these short-range triplet components \cite{You2004}.
The same mechanism is responsible for the sensitivity of $T_{c}$ on the degree of inhomogeneity in the model of the rotating magnetization \cite{Cha2004}. 

In the next section of this paper,  we develop our general approach 
within the framework of the quasiclassical theory of superconductivity.
We derive the  conditions required for the presence of long-range triplet components.
In Sec. III we investigate how the S/F proximity effect depends quantitatively on the degree of moment inhomogeneity in an S/F bilayer structure within the spiral order model. We provide the details of the calculations leading to the results already presented in our short paper \cite{Cha2004}.
In addition to Ref. \onlinecite{Cha2004}, we present the results concerning the spatial dependences of the singlet and short-range triplet components and discuss quantitatively the conditions which may increase or reduce the triplet amplitude.

\section{Spin structure of the Green functions induced by the S/F proximity effect}

In this section,
 we show that it is possible 
to capture general features of the S/F proximity effect independent of the specific geometry
 of the system under consideration.
For example, a spin splitting in the local density of states \cite{Faz1999} is  straightforwardly found to 
result from the production of the triplet components near the S/F interface \cite{Cha2004}. 
This spin imbalance
is obviously accompanied by the penetration of a spin magnetization in the superconductor \cite{Ber2004}. 
We shall also put forward the conditions and the physical reasons for the production of the long-range triplet components.

\subsection{Equations in Nambu-Gor'kov space}
We discuss the S/F proximity effect within the framework 
of the quasiclassical theory of superconductivity within the Nambu-Gor'kov formalism.
The particle-hole quantum-mechanical coherence is expressed by a
$2 \times 2$ matrix structure of the Green function in Nambu-Gor'kov space,
\begin{equation}
\hat{g}=\left( \begin{array}{cc}
g & f\\
\tilde{f} & \tilde{g}
\end{array}
\right).
\end{equation}
The functions $\tilde{g}$ and $\tilde{f}$ are the particle-hole 
conjugates of the Green functions $g$ and $f$.
We assume that the S/F bilayer we consider is in the diffusive limit,
in which case the Green function 
$\hat{g}({\bf R}, \omega_{n})$ depends on
a spatial coordinate ${\bf R}$, and on
the Matsubara frequencies $\omega_{n}=\pi T (2n+1)$ ($n$ an integer, $T$ the temperature).
The Green function obeys the 
Usadel transport equation for diffusive systems,\cite{Usa1970}
\begin{equation}
\left[i\, \omega_{n} \hat{\tau}_{3} - \hat{\Delta}-{\bf J}({\bf R}) \cdot \hat{\sigma}, \, \hat{g} \right]+ 
\frac{D_{ij}}{\pi} \nabla_{i}\left(  \hat{g} \nabla_{j}   \hat{g}\right)=0 \label{Usa1}
\end{equation}
with the normalization condition
\begin{equation}
\hat{g}^{2}=- \pi^{2} \hat{\tau}_{0} \label{No}
\end{equation}
where $D_{ij}$ is the diffusion constant tensor.
In Eq. (\ref{Usa1}),
we used the Einstein convention for summation over repeated indices 
($i,j=x$, $y$, $z$).  
We assume in our calculations for simplicity 
isotropic diffusion tensors, $D_{ij}=D \delta_{ij}$.
The diffusion constant $D$ in the superconductor differs a priori from the diffusion constant in the ferromagnet (an index characterizing both values is introduced below). We assume the same diffusion constants for both spin projections in the ferromagnet.
The superconducting order parameter $\hat{\Delta}$
and the spin matrix $\hat{\sigma}$ have the form
\begin{equation}
\hat{\Delta}=\left( \begin{array}{cc}
0& \Delta \\
\tilde{\Delta} &  0
\end{array}
\right) \hspace{0.2cm} \mathrm{and} \hspace{0.2cm}
\hat{\sigma}=\left( \begin{array}{cc}
\bm{\sigma} & 0 \\
0 &  \bm{\sigma}^{\ast}
\end{array}
\right).
\end{equation}
Here $\bm{\sigma}$ is the vector of spin Pauli matrices. We consider a superconductor with conventional s-wave pairing, so that $\tilde{\Delta}=\Delta^{\ast}$.
In the ferromagnet, the superconducting order parameter vanishes, $\Delta=0$. 
The proximity effect manifests itself in a nonzero pair amplitude $f \neq 0$ in the ferromagnet. 
In the superconductor, the exchange field vanishes, ${\bf J}\left({\bf R}\right)={\bf 0}$, 
expressing the non-coexistence of superconducting and ferromagnetic orders.
A non-zero magnetization $\delta {\bf M}({\bf R})$ in the superconductor may be induced
by the inverse proximity effect.

The Nambu-Gor'kov representation includes some redundancy which manifests itself in 
the fundamental symmetry relations\cite{Ser1983}
$\tilde{g}({\bf R},\omega_{n})=g^{\ast}({\bf R},\omega_{n})$,
$\tilde{f}({\bf R},\omega_{n})=f^{\ast}({\bf R},\omega_{n})$, and
\begin{eqnarray}
g({\bf R}, -\omega_{n}) &=&g^{\dag}({\bf R}, \omega_{n}) \label{symA} \\
f({\bf R}, -\omega_{n})& =&-f^{\mathrm{tr}}({\bf R}, \omega_{n}). \label{symB}
\end{eqnarray}
Here $g^{\dag}$ and $g^{\mathrm{tr}}$ are respectively the adjoint and transposed matrices.

\subsection{Spin structure}
The Green function $g$ and the anomalous Green function $f$ are $2 \times 2$ spin matrices.
For example, the anomalous Green function is written as
\begin{equation}
f=\left( \begin{array}{cc}
f_{\uparrow \uparrow} & f_{\uparrow \downarrow} \\
f_{\downarrow \uparrow}&  f_{\downarrow \downarrow}
\end{array}
\right).
\end{equation}
The order parameter in spin space for singlet pairing reads
\begin{equation}
\Delta= \Delta_{s} i \sigma_{y}.
\end{equation}

Generally $f$
can be written in the form (see e.g. Ref. \onlinecite{Ser1983,Ale1985,Tok1988,Min1999})
\begin{equation}
f= \left[f_{s}  + \left({\bf f}_{t}\cdot \bm{\sigma}\right) \right]i\sigma_{y} =\left( \begin{array}{cc}
-f_{tx}+if_{ty} & f_{tz}+f_{s} \\
f_{tz}-f_{s} & f_{tx}+if_{ty} \label{form1}
\end{array}
\right)
\end{equation}
where ${\bf f}_{t}=(f_{tx},f_{ty},f_{tz})$ is the triplet pairing vector and $f_{s}$ is the singlet pair amplitude. 
${\bf J}$ defines the spin-quantization axis (while in Sec. II we choose to coincide with the $z$-axis).
Thus, 
the probability to find a pair in the triplet state with zero spin projection on the quantisation axis $\hat{{\bf z}}$ is proportional to $\left|f_{tz} \right|^{2}$. When the vector ${\bf f}_{t}$ is directed along  the exchange field ${\bf J}$ (i.e. $f_{ty}=f_{tx}=0$), the Cooper pair spin is perpendicular to ${\bf J}$ and does not contribute to the spin paramagnetic susceptibility. 
When the triplet vector ${\bf f}_{t}$ is noncollinear with ${\bf J}$, it means that the amplitudes $\pm f_{tx}+if_{ty}$ of the states with spin projections $\pm 1$ on the quantization axis are non-zero. In this case, the Cooper pairs contribute to the susceptibility.

As for the Green function $g$, we adopt the notation
\begin{equation}
g=g_{0} + {\bf g}\cdot \bm{\sigma} =\left( \begin{array}{cc}
g_{0}+g_{z} & g_{x}-i g_{y} \\
g_{x}+i g_{y} & g_{0}-g_{z}
\end{array}
\right) \label{form2}
\end{equation}
with ${\bf g}=(g_{x},g_{y},g_{z})$ the vector part of $g$, and $g_{0}$ its scalar part.
Also, we expand the particle-hole conjugates in the spin space as
\begin{equation}
\tilde{f}=i \sigma_{y} \left[\tilde{f}_{s}- (\tilde{{\bf f}}_{t} \cdot \bm{\sigma} ) \right], \hspace*{0.2cm} \mathrm{and} \hspace*{0.3cm} \tilde{g}=\tilde{g}_{0} - \sigma_{y} \tilde{{\bf g}} \cdot \bm{\sigma} \sigma_{y} . \label{form3}
\end{equation}

\subsection{Normalization condition in spin-space}

The normalization condition (\ref{No}) written for the different components of the Nambu space yields the relations 
\begin{eqnarray}
gf+f \tilde{g}&=&0 \label{sym}\\
g^{2}+f \tilde{f}&=& - \pi^{2} \sigma_{0} \label{norm}
\end{eqnarray}
where $\sigma_{0}$ is the unit matrix in the spin-space.
In equilibrium, it can be shown from Eqs. (\ref{sym}) and (\ref{norm})
that for physical solutions necessarily the condition
\begin{equation}
\mathrm{Tr }\left(\hat{g} \right)=0,  \hspace*{0.4cm} \mathrm{i.e.} \hspace*{0.4cm} \tilde{g}_{0}=-g_{0} \label{hs}
\end{equation}
holds,
which expresses the particle-hole symmetry in the hybrid S/F structure.

Using the fact that the matrices $(\sigma_{0},\sigma_{x},\sigma_{y},\sigma_{z})$ form a basis for the spin matrices, we find from Eqs. (\ref{sym}) and  (\ref{hs}) that the different components of $f$ and $g$ in spin space obey the following conditions
\begin{eqnarray}
{\bf f}_{t} \cdot \left( {\bf g}-\tilde{{\bf g}}\right)&=& 0, \label{sym2a} \\
f_{s} \left( {\bf g}-\tilde{{\bf g}}\right)&=&i \, {\bf f}_{t} \times  \left( {\bf g}+\tilde{{\bf g}}\right) \label{sym2}.
\end{eqnarray}
Eq. (\ref{norm}) 
yields
\begin{eqnarray}
g_{0}^{2}+{\bf g}^{2}-f_{s} \tilde{f}_{s}+{\bf f}_{t} \cdot \tilde{{\bf f}}_{t}&=& - \pi^{2} \label{norm2}\\
2 g_{0} {\bf g}+f_{s} \tilde{{\bf f}}_{t}-\tilde{f}_{s} {\bf f}_{t}+i \, {\bf f}_{t} \times \tilde{{\bf f}}_{t}&=& {\bf 0} \label{last}
\end{eqnarray}
with ${\bf g}^{2}=g_{x}^{2}+g_{y}^{2}+g_{z}^{2}={\bf g} \cdot {\bf g}$ ($\neq \left|{\bf g}\right|^{2}$ if the components of ${\bf g}$ are complex numbers).
By combining the last equation (\ref{last}) with its particle-hole conjugate, we obtain  the relations
\begin{eqnarray}
g_{0} ({\bf g}-\tilde{{\bf g}})&=& i \, \tilde{{\bf f}}_{t} \times {\bf f}_{t} \label{rel1}\\
g_{0} ({\bf g}+\tilde{{\bf g}})&=&
\tilde{f}_{s} {\bf f}_{t}-f_{s} \tilde{{\bf f}}_{t} \label{rel2}.
\end{eqnarray}
As Eqs. (\ref{sym2a}) and (\ref{sym2}) follow from (\ref{rel1}) and (\ref{rel2}),
the normalization condition (\ref{No}) together with (\ref{hs})
leads to 3 independent equations (\ref{norm2}), (\ref{rel1}) and (\ref{rel2}) in spin-space.
Note that from Eq. (\ref{rel1}) and (\ref{rel2}) follows
${\bf g}^{2}= \tilde{{\bf g}}^{2}.$ However, in general  $\tilde{{\bf g}} \neq  {\bf g}$. According to Eq. (\ref{rel1}), the equality holds when ${\bf f}_{t} \parallel \tilde{{\bf f}}_{t}$.

Moreover, 
we see that  necessarily ${\bf g}=\tilde{{\bf g}}={\bf 0}$ 
when ${\bf f}_{t}={\bf 0}$. This implies a spin-independent density of states in the ferromagnet in the absence of the proximity effect.
This finding reflects the fact, that changes in the normal-state 
density of states as a result of an exchange splitting are small in
the expansion parameters of quasiclassical theory as long as the exchange splitting
is small compared to the conduction band widths for both spin directions.
Concordantly, we assume spin-independent diffusion constants for consistency.
In the opposite case, when the splitting is large compared to all low-energy
scales of the problem ($T$, $\Delta_s$), the spin splitting must be taken into
account as spin-dependent dispersions before applying the quasiclassical
approximation (see Ref. \onlinecite{Kop2004}).

\subsection{Quasi-classical transport equations in spin-space}
We now derive the transport equations for the spin components of $f$ and $g$.
For the component $f$ the equation (\ref{Usa1}) yields
\begin{eqnarray}
\frac{D}{\pi} \nabla_{j} \left(
g \nabla_{j} f+ f \nabla_{j} \tilde{g}
\right)+2 i \, \omega_{n} \, f
 =
\Delta \tilde{g} -g \Delta +\nonumber \\ 
+{\bf J} \cdot \bm{\sigma} \, f -f \, {\bf J} \cdot \bm{\sigma}^{\ast}.
\end{eqnarray}

The system of equations in the spin space reads (taking into account that $\tilde{g}_{0}=-g_{0}$)

\begin{eqnarray}
& & \frac{D}{\pi} \nabla_{j}
\left[g_{0} \nabla_{j} f_{s} - f_{s}\nabla_{j} g_{0} +{\bf g} \cdot \nabla_{j} {\bf f}_{t}-{\bf f}_{t} \cdot \nabla_{j} \tilde{{\bf g}}
\right]\nonumber \\
& & \qquad \qquad \qquad \qquad +2 i \, \omega_{n} \, f_{s} = 
-2 g_{0}\Delta_{s} + 2 \, {\bf J} \cdot \bf{f}_{t} \qquad
\label{Usas}
\\ 
& & \frac{D}{\pi} 
\nabla_{j} \left[
g_{0} \nabla_{j} {\bf f}_{t}-{\bf f}_{t} \nabla_{j} g_{0}+ {\bf g} 
\nabla_{j} f_{s}-f_{s}\nabla_{j} \tilde{{\bf g}} +i \, {\bf g} \times \nabla_{j} {\bf f}_{t}\right.
\nonumber \\ 
& & \quad \left. +i \, \nabla_{j} \tilde{{\bf g}} \times {\bf f}_{t}
\right]+
2 i \, \omega_{n} \, {\bf f}_{t} = 
-\Delta_{s}\left(\tilde{{\bf g}}+{\bf g}\right) + 2 \, {\bf J}f_{s}. \qquad
\label{Usat}
\end{eqnarray}

It is supplemented by the self-consistent equation defining the s-wave order parameter $\Delta$ from $f_{s}$ which reads in the dirty limit for a weak-coupling 
\begin{equation}
\Delta_{s}({\bf R})=\lambda T \sum_{n} f_{s}({\bf R},\omega_{n}) \label{gap}
\end{equation}
with $\lambda$ the pairing interaction constant.
Using the equation
\begin{eqnarray}
\pi T \sum_{n}
\frac{1}{\left|\omega_{n}\right|}=\frac{1}{\lambda}+\ln \frac{T_{c0}}{T} \nonumber
\end{eqnarray}
relating the critical temperature $T_{c0}$ of the superconductor without the proximity of a ferromagnetic layer and $\lambda$,
 we can rewrite the Eq. (\ref{gap}) as
\begin{equation}
\Delta_{s}\ln \frac{T_{c0}}{T}=\pi T \sum_{n} \left(
\frac{\Delta_{s}}{\omega_{n}}-
\frac{f_{s}(\omega_{n})}{\pi} \label{self}
\right).
\end{equation}

Near the critical temperature $T_{c}$, 
the pair amplitudes $f_{s}$ and ${\bf f}_{t}$ are small and the Green function $g$ deviates only slightly from its value ($g=-i \pi  \sigma_{0} \, \mathrm{sgn} \, \omega_{n}$) in the normal state, so that the Usadel equations (\ref{Usas})-(\ref{Usat}) can be linearized and take the simpler form \cite{Cha2004}

\begin{eqnarray}
\left(D \bm{\nabla}^{2}-2|\omega_{n}| \right)&f_{s}=& -2\pi \Delta_{s}+2i\, \mathrm{sgn}(\omega_{n}) \, {\bf J} \cdot {\bf f}_{t} \label{fs}
\\
\left(D \bm{\nabla}^{2}-2|\omega_{n}| \right)&{\bf f}_{t}=& 2i \, \mathrm{sgn}(\omega_{n}) \,
{\bf J}f_{s}. \label{coup}
\end{eqnarray}

\subsection{General features of the S/F proximity effect}

Within a particular geometry for the S/F hybrid structures, we have to complement the transport equations with boundary conditions.
Nevertheless, many general features of the S/F proximity effect such as the number and the nature of the nonzero spin components for $f$ and $g$  can be determined independently from the geometric effects of the boundaries.

From the transport Eq. 
(\ref{Usat}), and even clearer from Eq. (\ref{coup}), it follows that the triplet vector ${\bf f}_{t}$ is necessarily non-zero if the singlet component $f_{s}$ penetrates in the ferromagnet.
Moreover,  
the triplet vector tends to align with the exchange field ${\bf J}$, indicating
that the dominant triplet component has
zero spin-projection. 
Thus,  the singlet component and the triplet component with zero spin-projection coexist always in the ferromagnet near the S/F interface.
This is physically expected since these two pair correlations 
 are energetically equivalent in the ferromagnet with respect to their interaction  with the exchange field.
Both components are characterized by short-range penetration lengths in the ferromagnet. 

On the contrary,  if the triplet vector ${\bf f}_{t}$ is non-collinear with ${\bf J}$, it means that triplet components with nonzero spin-projection on ${\bf J}$ are produced. Since these correspond to equal-spin pairing, they are not limited locally by the paramagnetic interaction with the local exchange field and may
have long-range scales in the ferromagnet.
A misalignment between the triplet vector ${\bf f}_{t}$ and the moment ${\bf J}$ occurs in presence of sudden changes in orientation of ${\bf J}$. The reason is that ${\bf f}_{t}$ obeys a differential equation and its variations in orientation have thus to be relatively smooth.

From Eq. (\ref{rel2}) it is clear that the 
counterpart for the production of triplet components (${\bf f}_{t} \neq {\bf 0}$) for $f$ is the presence of ${\bf g} \neq {\bf 0}$ for $g$.
As a direct consequence, the density of states for the up and down spin projections differs.\cite{Faz1999}
Furthermore,
in the presence of long-range triplet components, the Green function $g$ contains also off-diagonal (spin-flip) terms.
As a result of the spin splitting in the density of states  generated by the S/F proximity effect, 
a spin magnetization $\delta {\bf M}$ is induced near the S/F interface. 
This magnetization leakage has been investigated recently in the Ref. \onlinecite{Ber2004} within a model considering a fixed exchange field.  
The spin magnetization induced by the proximity effect is given by \cite{Ale1985,Tok1988}
\begin{equation}
\delta {\bf M}({\bf R})=2 N_{0} T \sum_{n} {\bf g}\left({\bf R}, \omega_{n}\right) \label{aim}.
\end{equation}
Since the triplet vector ${\bf f}_{t}$ is also induced in the superconductor near the S/F interface  via an inverse proximity effect,
the vector ${\bf g}$ characterizing the magnetic correlations penetrates also in the superconductor according to the relation (\ref{rel2}). 

As we show in the following, the sum over Matsubara frequencies in the expression (\ref{aim}) is in general nonzero.
Indeed, as noticed in the papers \cite{Ber2001,Vol2003,Ber2003}, in the diffusive
limit the triplet components have the property to be odd functions of the Matsubara frequencies $\omega_{n}$, while the singlet amplitude $f_{s}$ is an even function of $\omega_{n}$. These properties  follow directly from the Pauli
principle, which  leads to the
relation (\ref{symB}). The different components of  $g$ have also
symmetry properties with respect to $\omega_{n}$. 
Combining (\ref{hs}) with (\ref{symA}),
we obtain that $g_{0}$ is an odd function of $\omega_{n}$, and
\begin{eqnarray}
{\bf g}(-\omega_{n})=\tilde{{\bf g}}(\omega_{n}). \label{dirty}
\end{eqnarray}

We note that the relations (\ref{rel1})-(\ref{rel2}) between the vectors ${\bf g}$ and ${\bf f}_{t}$ are simplified
when ${\bf f}_{t} \times \tilde{{\bf f}}_{t}={\bf 0}$.
This condition corresponds to unitary triplet superconductivity \cite{Min1999}.
It can be seen from the transport equations (\ref{Usas})-(\ref{Usat}) that 
 if the gap $\Delta_{s}$ can be chosen real (and thus the singlet amplitude $f_{s}$ is real), then the triplet vector ${\bf f}_{t}$ is purely imaginary, i.e. $\tilde{{\bf f}}_{t}={\bf f}_{t}^{\ast}=-{\bf f}_{t}$
(taking into account that $\tilde{g}_{0}=-g_{0}$, i.e. $g_{0}$ is purely imaginary). As a result, one obtains the simpler relations
\begin{equation}
{\bf g}=\tilde{{\bf g}}, \hspace*{1cm} g_{0} {\bf g}= f_{s} {\bf f}_{t}. \label{simp}
\end{equation}
These simplifications are found also if ${\bf f}_{t} \parallel {\bf J}$ (i.e. when no long-range triplet components are induced). Combining the relations (\ref{simp}) and (\ref{dirty}), we conclude that the spin-vector part ${\bf g}$ is an even function of $\omega_{n}$ under certain circumstances, which demonstrates that the induced spin magnetization $\delta {\bf M}$ is non-zero.

Therefore, although there is no coexistence of ferromagnetic and superconducting orders, magnetic and superconducting correlations do coexist in the vicinity of  both sides of the S/F interface. 
It is worth noting that near $T_{c}$, the singlet amplitude $f_{s}$ and the triplet vector amplitude are small so that  ${\bf g}$ and thus $\delta {\bf M}$ appear to be second order terms (see Eq. (\ref{rel2})).
Accordingly, one expects that  the induced spin magnetization $\delta {\bf M}$ penetrating the superconductor, which is negligibly small near $T_{c}$, increases significantly by reaching temperatures well below $T_{c}$.

\section{S/F bilayer with a rotating exchange field}

In this section, we study the proximity effect in a S/F bilayer within the model of a rotating exchange field in F (see Fig.~\ref{Fig1}).
We derive analytical expressions for the spatial dependences of the singlet and triplet amplitudes near the superconducting critical $T_{c}$ as a function of the spiral wavevector $Q$. We calculate also numerically the dependence of $T_{c}$ on $Q$.
A part of the results has been already reported in our short paper \cite{Cha2004}. 
We shall see here that the magnetization induced in the superconductor by the triplet components reflects the inhomogeneity of the exchange field ${\bf J}$ in the ferromagnet. 

\subsection{Model and boundary conditions}

In the F layer (defined by the plane $(x,y)$),  ${\bf J}$ rotates with an angle varying along the direction $y$, i.e. 
\begin{equation}
{\bf J}(y)=J ( \cos Q y, \sin Qy,0).
\end{equation}
It is straightforward to see that in the present geometry the component $f_{tz}$ of the triplet vector is zero (because ${\bf J}$ never points in this direction).  
It is worth mentioning  that here $z$ does not correspond to the 
spin quantization axis. To identify physically the different components of the triplet vector, one has to express ${\bf f}_{t}$ in the local basis $(X,Y,Z)$ where $Z$ is the direction of the local exchange field ${\bf J}$.
According to the symmetry relations (see the previous section),
it is sufficient to consider the positive Matsubara frequencies.
We want to determine the  dependence of the critical temperature $T_{c}$ on the rotation wavevector $Q$, and for this purpose we have to solve the linearized Usadel equations.
The $y$ dependence of the moment ${\bf J}$ is eliminated in the right-hand side of the Eq. (\ref{fs})-(\ref{coup}) by considering the new components 
\begin{eqnarray} 
f_{+}&=&(-f_{tx}+if_{ty}) \, e^{i Qy} \\
f_{-}&=&(f_{tx}+if_{ty})\, e^{-i Qy}.
\end{eqnarray} 
The new system of equations to solve takes the form
\begin{eqnarray}
\left(D\bm{\nabla}^{2} -2\omega_{n} \right)
f_{s}= -2 \pi \Delta_{s}+
 i \, J \left(f_{-}-f_{+}\right)  \label{syst1}
\\
\left(D \bm{\nabla}^{2} \mp 2 i DQ \partial_{y}
-DQ^{2} -2\omega_{n}\right)f_{\pm}= \mp \, 2 i \, Jf_{s} 
. \label{syst2}
\end{eqnarray}
The diffusion constants are $D=D_{s}$ in the S layer and $D=D_{f}$ in F.

The components of the triplet vector in the plane $(x,y)$ are obtained from $f_{+}$ and $f_{-}$ with the relations
\begin{eqnarray}
f_{tx}=\frac{1}{2}\left(f_{-}e^{iQy}-f_{+}e^{-iQy} \right), \label{trip1}\\
f_{ty}=\frac{1}{2i}\left(f_{-}e^{iQy}+f_{+}e^{-iQy} \label{trip2}
\right).
\end{eqnarray}
In the present case, the gap amplitude $\Delta_{s}$ and the singlet amplitude $f_{s}$ can be chosen real. Then $f_{-}=f_{+}^{\ast}$, i.e.
\begin{eqnarray}
f_{tx}=i \left( - \mathrm{Im}\, f_{+} \, \cos Qy 
+\mathrm{Re} \,f_{+} \, \sin Qy 
\right) \label{fx}\\
f_{ty}=-i \left( \mathrm{Re}\, f_{+} \, \cos Qy 
+\mathrm{Im} \,f_{+} \, \sin Qy 
\right). \label{fy}
\end{eqnarray}
It is clear that the amplitudes of the triplet components are purely imaginary.

The system of equations (\ref{syst1})-(\ref{syst2}) is supplemented by boundary conditions. The boundary conditions  at the S/F interface 
(provided that $J \ll \varepsilon_{F}$ with $\varepsilon_{F}$ the Fermi energy)
for the diffusive regime have been formulated by Kuprianov and Lukichev for a small barrier transparency 
\cite{Kup1988}.
The general boundary conditions  have been derived by Nazarov \cite{Naz1999}. 
Near $T_{c}$, they are formally equivalent and reduce to

\begin{equation}
\xi_{s} \left. \partial_{z} f \right)_{SC}=
\gamma \xi_{f} \left. \partial_{z} f  \right)_{F}, \hspace{0.5cm} \gamma=\rho_{s} \xi_{s}/\rho_{f}\xi_{f} \label{bc1}
\end{equation}
where 
$\rho_{s}$ and $\rho_{f}$ are respectively the normal-state resistivities of the S and F metals
(this boundary condition follows from the continuity of the current at the interface), and
\begin{equation}
\xi_{f} \gamma_{b} \left. \partial_{z} f \right)_{F}= \left. f \right)_{SC}-\left. f \right)_{F}, \hspace{0.5cm} \gamma_{b}=R_{b} {\cal A}/\rho_{f} \xi_{f} \label{bc2}
\end{equation}
with $R_{b}$ the resistance of the S/F boundary, and ${\cal A}$ its area. 
Here $\xi_{s}=\sqrt{D_{s}/2 \pi T_{c0}}$ is the superconducting coherence length in S and $\xi_{f}=\sqrt{D_{f}/2 \pi T_{c0}}$ is the superconducting coherence length in F.
At the outer surfaces of the F or S layers ($z=-d_{f}$ and $z=d_{s}$), the current through the boundary has to vanish, i.e.
\begin{equation}
\partial_{z} f =0. \label{bc3}
\end{equation}
It is important to note that 
 the present boundary conditions (\ref{bc1})-(\ref{bc3}) do not couple the different spin components  of $f$. 

\subsection{In the ferromagnet}

In the ferromagnetic layer, the singlet amplitude and the triplet components $f_{\pm}$ are coupled through
\begin{eqnarray}
\left(\bm{\nabla}^{2} -\Omega_{n} \xi_{f}^{-2} \right)
f_{s}=
 i \, \xi_{J}^{-2}\left(f_{-}-f_{+}\right) \label{F1}
\\
\left(\bm{\nabla}^{2} \mp 2 i Q \partial_{y}
-Q^{2} -\Omega_{n} \xi_{f}^{-2} \right)f_{\pm}= \mp \, 2 i \, \xi_{J}^{-2}f_{s} \label{F2}
.
\end{eqnarray}
where $\xi_{J}=\sqrt{D_{f}/J}$
and $\Omega_{n}=(2 n +1)T/T_{c0}$.

Since the geometry is periodic in the $y$ direction,
the components of the superconducting condensate wave function $f$ can be 
expanded into a Fourier series. 
Using  the boundary condition at the outer surface ($z=-d_{f}$), the components of $f$ are sought 
 under the form
\begin{equation}
f_{l}(y,z)= \sum_{p=-\infty}^{+\infty} f_{l}^{(p)} \cosh\left[k_{fp} (z+d_{f}) \right] e^{ipQ y}
\end{equation}
where $l=s,\pm$. 
Substituting these expressions in the set of equations (\ref{F1})-(\ref{F2}) leads to the following linear system 
\begin{equation}
\left(\begin{array}{ccc}
\tilde{k}_{p}^{2}-Q^{2}_{p} & -i \xi_{J}^{-2} & i \xi_{J}^{-2}  \\
-2i \xi_{J}^{-2} & \tilde{k}_{p}^{2}-Q^{2}_{p+1}& 0 \\
2 i \xi_{J}^{-2} & 0& \tilde{k}_{p}^{2}-Q^{2}_{p-1}
\end{array}
\right)
\left(\begin{array}{c}
f_{s}^{(p)}\\
f_{-}^{(p)}\\
f_{+}^{(p)}
\end{array}
\right)
=0
\end{equation}
with $\tilde{k}^{2}_{p}=k^{2}_{fp}-\Omega_{n}\xi_{f}^{-2}$ and $Q_{p}=pQ$.
The eigenvalues $k_{fp}^{2}$ are determined from the condition of zero determinant for the $3 \times 3$ matrix.
For $p \neq 0$ the three eigenvalues $k_{fpj}^{2}$ ($j=1$, $2$ and $3$) and the associated  eigenvectors $\left(f_{s,p,j},f_{-,p,j},f_{+,p,j}\right)$ take a complicated analytical form that we shall not give here.  
We note that for $p \neq 0$, the three components of each eigenvectors $f_{l,p,j}$ are non zero.

For $p=0$, the
 three eigenvalues can be easily derived and have a simple form
\begin{equation}
k^{2}_{f0\varepsilon}=\Omega_{n} \xi_{f}^{-2}+ \varepsilon \, 2 i \, \xi_{J}^{-2} \eta_{-\varepsilon}
\hspace{0.2cm} \mathrm{and} \hspace{0.2cm}
k^{2}_{f03}=\Omega_{n}\xi_{f}^{-2}+Q^{2} \label{value}
\end{equation}
 with $\eta_{\varepsilon}= \sqrt{1-\eta^{2}}+i \varepsilon \eta$ for $\eta \leq 1$, $\varepsilon=\pm 1$, $\eta=\xi_{J}^{2}Q^{2}/4$. For $\eta \geq 1$, we have  $\eta_{\varepsilon}= -i\sqrt{\eta^{2}-1}+i \varepsilon \eta$
. The two first eigenvalues (index $\varepsilon =+$ or equivalently $j=1$, and $\varepsilon=-$ or $j=2$) which are related by a complex conjugation for $\eta \leq 1$ correspond to a short penetration length (at least at small $Q$) of the order of $\xi_{J} \ll \xi_{f}$ for strong ferromagnets ($J \gg 2 \pi T_{c0}$). On the contrary, the third eigenvalue ($k_{f03}^{2}$), which is independent of the exchange field amplitude $J$, is real and gives a much longer decay length of the order of $\xi_{f}$ for the pair amplitude in the ferromagnet. 
The corresponding eigenvectors have the form
\begin{equation}
\left(
\begin{array}{c}
f_{s,\varepsilon}\\
f_{-,\varepsilon}\\
f_{+ ,\varepsilon}
\end{array}
\right)=
\left(
\begin{array}{c}
\eta_{\varepsilon}
\\
\varepsilon\\
- \varepsilon
\end{array}
\right)
\hspace{0.1cm}
\mathrm{and}
\hspace{0.1cm}
\left(
\begin{array}{c}
f_{s,3}\\
f_{-,3}\\
f_{+,3}
\end{array}
\right)= 
\left(
\begin{array}{c}
0\\
1\\
1
\end{array}
\right). \label{eigen}
\end{equation}
In the limit $Q \to 0$, then $\eta \to 0$ and $\eta_{\varepsilon} \to 1$, and we find again from (\ref{value}) the known eigenvalues for a fixed exchange field \cite{Rad1991,Ber2001}. 

The general solution of the system (\ref{F1})-(\ref{F2}) satisfying the outer boundary condition can be written as 
\begin{equation}
f_{l}(y,z)= \sum_{p=-\infty}^{+\infty}\sum_{j=1}^{3} a_{j}^{(p)} f_{l,p,j} \cosh\left[k_{fpj} (z+d_{f}) \right] e^{ipQ y}
\end{equation}
where the three coefficients $a_{j}^{(p)}$  have to be determined with the help of the boundary conditions at the S/F interface.

\subsection{In the superconductor}

\subsubsection{Triplet components}

In the S layer, there is no coupling by the equations between the singlet and triplet components.
The solutions for $f_{\pm}$ satisfying the boundary condition at the outer surface (at $z=d_{s}$) and being periodical in the $y$ direction are straightforwardly derived and have the form
 \begin{equation}
f_{\pm}(y,z)= \sum_{p=-\infty}^{+\infty}f_{\pm}^{(p)}
\cosh\left[k_{tp}^{\pm}(z-d_{s})\right] \, e^{ipQy}
\end{equation}
where 
\begin{equation}
k_{tp}^{\pm}=\sqrt{\Omega_{n} \xi_{s}^{-2}+Q_{p \mp 1}^{2}}
\end{equation}
and the coefficients $f_{\pm}^{(p)}$  have to be determined with the boundary conditions at the S/F interface.

\subsubsection{Singlet component}

In the S layer, the equation for the singlet pair amplitude $f_{s}$ 
\begin{equation}
\left(D_{s} \bm{\nabla}^{2} -2\omega_{n} \right)
f_{s}=- 2 \pi  \Delta_{s}
\end{equation}
is coupled to the self-consistency equation (\ref{self}).
Due to the periodic geometry in the $y$ direction, $f_{s}$ and $\Delta_{s}$ can also be expanded into Fourier series
\begin{eqnarray}
f_{s}(y,z)=\sum_{p=-\infty}^{+\infty}f_{s}^{(p)}(z)
\, e^{ipQy}
\\
 \Delta_{s}(y,z)=\sum_{p=-\infty}^{+\infty}\Delta_{s}^{(p)}(z)
\, e^{ipQy}.
\end{eqnarray}
Then, the Fourier amplitude $f_{s}^{(p)}(z)$ obeys the differential equation
\begin{equation}
\left(\partial_{z}^{2}-Q_{p}^{2}-\Omega_{n} \xi_{s}^{-2}\right)
f_{s}^{(p)}=- 2 \pi  \Delta_{s}^{(p)}/D_{s} \label{equat1}
\end{equation}
while the gap amplitude  $\Delta_{s}^{(p)}(z)$  is given by
\begin{equation}
\Delta_{s}^{(p)}(z)\ln \frac{T_{c0}}{T}=2\pi T \sum_{n \geq 0} \left(
\frac{\Delta_{s}^{(p)}(z)}{\omega_{n}}-
\frac{f_{s}^{(p)}(z)}{\pi}
\right). \label{gap2}
\end{equation}
In general, the coupled equations (\ref{equat1})-(\ref{gap2}) can not be solved analytically due to the self-consistence.

It is clear at this point that there is no mixing between the different Fourier components of the superconducting condensate function $f$ neither in the ferromagnet nor in the superconductor near the critical temperature.
It means that each  single Fourier component $p$ (which determines  a particular $y$ dependence) is a solution of the system of equations which satisfies the boundary conditions.
For each $p$, we obtain a different gap equation (\ref{gap2}), i.e. a different critical temperature $T_{c}(p)$.
The solution realized physically is the one which gives the highest critical temperature (i.e. which is energetically most favorable).

For $p \neq 0$, the singlet amplitude $f_{s}$ and the order parameter $\Delta_{s}$ are inhomogeneous along the direction $y$ even far from the S/F interface.
The solution with $p=0$ corresponds to a singlet amplitude $f_{s}$ which depends only on the spatial variable $z$ both in the F and S layers. In this case, the components $f_{\pm}$ are also independent of the coordinate $y$ (characterizing the spatial inhomogeneity of ${\bf J}$). The $y$ dependence of the triplet vector is then simply revealed by the relations (\ref{trip1})-(\ref{trip2}). The influence of the inhomogeneity of the exchange field on the singlet amplitude $f_{s}$ in the superconductor occurs near $T_{c}$ only through the boundary conditions at the S/F interface.

\subsection{Pair amplitudes}
The remaining  step is to determine the coefficients $a_{j}^{(p)}$ and $f_{\pm}^{(p)}$ with the boundary conditions at the S/F interface. Let us define the short-hand notation $\delta_{0}^{(p)}=f_{s}^{(p)}(z=0)$.
The condition (\ref{bc2}) yields
\begin{eqnarray}
\delta_{0}^{(p)}= \sum_{j=1}^{3} a_{j}^{(p)} f_{s,p,j} {\cal A}_{j}^{(p)}  \label{s1}\\
f_{l}^{(p)} \cosh (k_{tp}^{l}d_{s})= \sum_{j=1}^{3} a_{j}^{(p)} f_{l,p,j} {\cal A}_{j}^{(p)} \label{deux}
\end{eqnarray}
where $l=\pm$ and
\begin{equation}
{\cal A}_{j}^{(p)}=\cosh (k_{fpj}d_{f})+\gamma_{b} k_{fpj} \xi_{f} \sinh (k_{fpj}d_{f}).
\end{equation}
Then, the condition (\ref{bc1}) considered for the triplet components only, combined with the Eq. (\ref{deux}) gives 
\begin{eqnarray}
\sum_{j=1}^{3} a_{j}^{(p)} f_{l,p,j} \tilde{{\cal A}}_{j,l}^{(p)} =0 \label{s2}
\end{eqnarray}
with 
\begin{eqnarray}
\tilde{{\cal A}}_{j,l}^{(p)}&=&{\cal A}_{j}^{(p)}+ C_{j}^{(p)} \,(k_{tp}^{l}\xi_{s}\tanh (k_{tp}^{l}d_{s}))^{-1} \\
C_{j}^{(p)}&=&\gamma k_{fpj} \xi_{f} \sinh (k_{fpj}d_{f})
.
\end{eqnarray}
The equations (\ref{s1}) and (\ref{s2}) lead to a $3 \times 3$ linear system for the coefficients $a_{j}^{(p)}$. As a result, we obtain straightforwardly for $p \neq 0$
\begin{eqnarray}
a_{j}^{(p)}=\delta_{0}^{(p)} \frac{N_{j}}{{\bf F} \cdot {\bf N}} \label{expr}
\end{eqnarray}
with ${\bf N}= {\bf h}_{+} \times {\bf h}_{-}$, and where ${\bf h}_{l}$  ($l=\pm$) and ${\bf F}$ are vectors whose components labeled by $j=1$, 2 and $3$ are given by
\begin{equation}
F_{j}= f_{s,p,j}  {\cal A}_{j}^{(p)}, \hspace*{1cm} h_{l,j}=f_{l,p,j} \tilde{{\cal A}}_{j,l}^{(p)}.
\end{equation}

For $p=0$, the absence of coupling between the singlet amplitude $f_{s,3}$ and the two triplet amplitudes $f_{\pm,3}$ for the third eigenvector in (\ref{eigen}) yields that the long-range contribution is absent ($a_{3}^{(0)}=0$). By the same occasion, the quantity $\tilde{{\cal A}}_{j,l}^{(0)}$ is independent of the index $l=\pm$ (which is dropped out in the forthcoming expressions).
According to the form of the eigenvectors associated with the eigenvalues $k_{f \varepsilon}^{2}$ in Eq. (\ref{eigen}) (from now on we drop 
the index indicating that $p=0$), the two triplet components $f_{+}$ and $f_{-}$ in the ferromagnet are related as $f_{+}=-f_{-}$. Since in addition the symmetry property $f_{+}=f_{-}^{\ast}$ holds, we have necessarily $\mathrm{Re} \, f_{+}=0$.
Then using the Eq. (\ref{fx})-(\ref{fy}), we find that the corresponding triplet vector has the spatial dependence
\begin{eqnarray}
f_{tx}(y,z)=i \, \mathrm{Im}\, f_{-}(z) \, \cos Qy \\
f_{ty}(y,z)=i \, \mathrm{Im} \,f_{-}(z) \, \sin Qy ,
\end{eqnarray}
i.e. ${\bf f}_{t}$ follows everywhere the direction of the inhomogeneous exchange field ${\bf J}(y)$. Thus, only the triplet vector  component $f_{tZ}$ with zero spin-projection on the local moment ${\bf J}$ (direction $Z$) exists. One obtains that 
\begin{equation}
f_{tZ}(z)=f_{-}(z)=\sum_{\varepsilon=\pm} a_{\varepsilon} \varepsilon \cosh\left[ k_{f \varepsilon} (z+d_{f})\right]
\end{equation}
is characterized by a short penetration length in the ferromagnet, as expected from the physical arguments presented in the Section II. In the F layer, $f_{tZ}$ exhibits  in addition oscillations in space as long as the eigenvalues $k_{f \varepsilon}^{2}$ are complex.
The two remaining nonzero amplitudes $a_{+}$ and $a_{-}$ in the ferromagnet (for convenience we label the amplitudes by $\varepsilon =\pm$ rather than  $j=1$, $2$) obey the 2 x 2 system determined by the Eq. (\ref{s1}) and (\ref{s2}), and are given by  
\begin{eqnarray}
a_{+}&=&\delta_{0}
\frac{\tilde{{\cal A}}_{2}}{\eta_{+} {\cal A}_{1} \tilde{{\cal A}}_{2}+ \eta_{-} {\cal A}_{2} \tilde{{\cal A}}_{1}}\\
a_{-}&=&\delta_{0}
\frac{\tilde{{\cal A}}_{1}}{\eta_{+} {\cal A}_{1} \tilde{{\cal A}}_{2}+ \eta_{-} {\cal A}_{2} \tilde{{\cal A}}_{1}}.
\end{eqnarray}
For $\eta \leq 1$, we have $a_{+}=a_{-}^{\ast}$, while for $\eta \geq 1$ the coefficents $a_{\pm}$ are purely imaginary.
We find from the Eq. (\ref{deux}) that the triplet component with zero spin projection has the following spatial dependence in the S layer
\begin{equation}
f_{tZ}(z)=\left(a_{+} {\cal A}_{1}-a_{-} {\cal A}_{2} \right) \frac{\cosh\left[ k_{t}(z-d_{s})\right]}{\cosh(k_{t}d_{s})}
\end{equation}
with the real wavevector $k_{t}=\sqrt{\Omega_{n} \xi_{s}^{-2}+Q^{2}}$.
Finally, it is straightforward to see that for $p=0$, the spin-vector part ${\bf g}$ (an even function of $\omega_{n}$ here) of the Green function $g$ determined from the singlet and triplet components according to the relation (\ref{simp}) can be written near $T_{c}$ as 
\begin{equation}
{\bf g}=\frac{i}{\pi}f_{s}(z) f_{tZ}(z) \, \hat{{\bf J}}(y).
\end{equation}
   From this equation, we note that the spin magnetization induced in the superconducting layer exhibits the same inhomogeneity as the exchange field ${\bf J}(y)$ in the transverse direction $y$.

\subsection{Superconducting critical temperature $T_{c}$}

Using the last boundary condition  (\ref{bc1}) considered for the singlet amplitude and the expression (\ref{expr}), we derive a relation between the derivative of the singlet component $f_{s}^{(p)}(z)$ and its value $\delta_{0}^{(p)}$ at the interface as

\begin{equation}
\xi_{s} \, f_{s}^{(p)'}(z=0)
=W_{p} \, \delta_{0}^{(p)} \label{W}
\end{equation}
with 
\begin{equation}
W_{p}=
\frac{{\bf C}
\cdot {\bf N}
}
{{\bf F}
\cdot {\bf N}} .
\end{equation}
For $p=0$, 
the function $W_{0}$ can be written as
\begin{equation}
W_{0}=
\frac{\eta_{+}C_{1} \tilde{{\cal A}}_{2}+ \eta_{-} C_{2} \tilde{{\cal A}}_{1}}{\eta_{+}{\cal A}_{1} \tilde{{\cal A}}_{2}+\eta_{-}{\cal A}_{2} \tilde{{\cal A}}_{1}}.
\end{equation}
In the limit $Q \to 0$, 
we recover from this expression the formula (12) of the Ref.~\onlinecite{Fom2002} obtained for a S/F bilayer with a constant exchange field ${\bf J}$ in the ferromagnetic layer. 
All the informations characterizing the proximity effect between the S and F layers  are contained in this real function $W_{p}$. 

 Following the Ref.~\onlinecite{Fom2002}, the solution of
the Eq. (\ref{equat1}) satisfying  the boundary conditions at the outer surface (Eq. (\ref{bc3}) for $z=d_{s}$) and at the S/F interface (Eq. (\ref{W}))
is expressed using the Green function of the equation. As a result, one can write
the singlet pair amplitude as
\begin{equation}
f_{s}^{(p)}(z)= \pi \int_{0}^{d_{s}} \!\!\!\!\! G(y,z) \Delta^{(p)}(y)dy
. \label{fsing}
\end{equation}
with the Green function $G$ given by

\begin{eqnarray}
G(y,z)=\frac{(k_{s}\xi_{s}^{2} \pi T_{c0})^{-1}}{\sinh(k_{s}d_{s})+(W_{p}/k_{s}\xi_{s}) \cosh(k_{s}d_{s})} \times \nonumber \\
\times \left\{
\begin{array}{cc}
v_{1}(z) \, v_{2}(y), & z \leq y \\
v_{2}(z) \, v_{1}(y), & y \leq z 
\end{array}
\right.
\end{eqnarray}
where 
\begin{eqnarray}
v_{1}(z)&=&\cosh(k_{s}z)+(W_{p}/k_{s}\xi_{s}) \sinh(k_{s}z),\\
v_{2}(z)&=&\cosh(k_{s}[z-d_{s}]),
\end{eqnarray}
and 
\begin{equation}
k_{s}=\sqrt{\Omega_{n} \xi_{s}^{-2}+Q_{p}^{2}}. \label{ks}
\end{equation}
Combining the equations (\ref{gap2}) and (\ref{fsing})-(\ref{ks}) one obtains a single equation to be solved numerically. 


\subsection{Numerical results and Discussion}

\begin{figure}[t]
\begin{center}
\includegraphics[height=6.5cm]{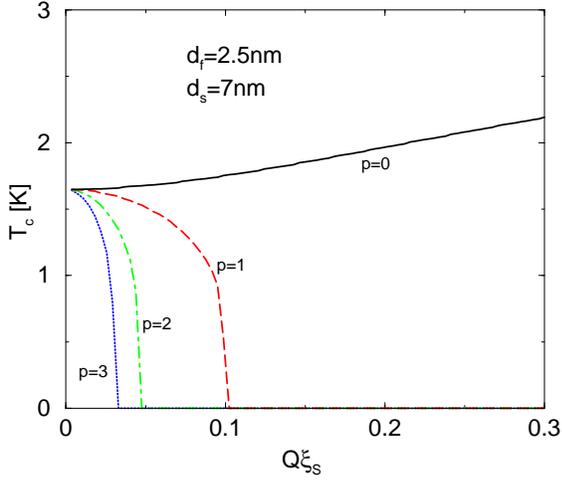}
\caption{$T_{c}$ versus the spiral wavevector $Q$ for different harmonic solutions. The highest critical temperature at $Q \neq 0$ is obtained for $p=0$. Here $\gamma_{b}=0.3$,  $d_{f}=2.5$ nm and $d_{s}=7$ nm.} \label{FigTcp}
\end{center}
\end{figure}

The dependence of $T_{c}$ on the spiral wavevector $Q$ for the different harmonic solutions is plotted in Fig.~\ref{FigTcp}
(for definiteness, we took here similar parameters as in Fig.~2 of Ref.~\onlinecite{Fom2002}, that is $T_{c0}=7$ K, $\gamma=0.15$, $J=130$ K, $\xi_{s}=8.9$ nm and $\xi_{f}=7.6$ nm). 
As shown in this figure, the harmonic
$p=0$ yields the highest $T_{c}$. 
We found the same result for a large range of input parameters (e.g. for bigger thicknesses $d_{s}$ and $d_{f}$, not shown here). The solutions with $p \neq 0$ correspond to a superconducting state in the S layer which remains inhomogeneous far from the S/F interface (corresponding to a
Fulde-Ferrell-Larkin-Ovchinnikov state\cite{Ful1964,Lar1965}).
They are energetically less favorable than the solution with $p=0$ which gives a singlet pair amplitude that is inhomogeneous only near the interface due to the proximity effect
(the fact that the singlet correlations are independent of the coordinate $y$ for a $y$-independent order parameter
simply reflects their isotropy in the spin space: singlet pairs do not see the directional changes of the exchange field).

\begin{figure}[t]
\begin{center}
\includegraphics[height=5.6cm]{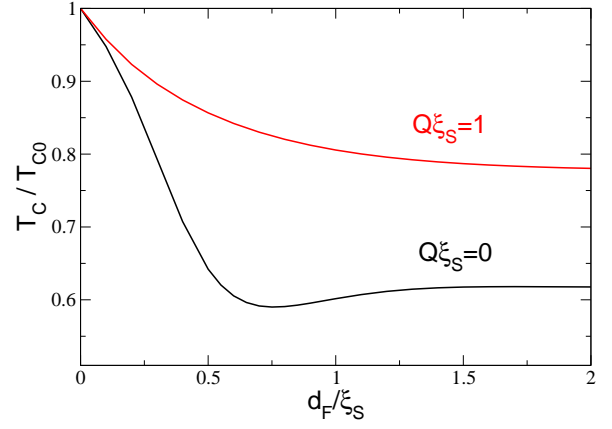}
\caption{$T_{c}$ versus  $d_{f}$ for $Q\xi_{s}=0$ and $Q \xi_{s}=1$. The characteristic non-monotonic dependence of $T_{c}(d_{f})$ found at $Q=0$ is suppressed by the moment inhomogeneity in F.  Here, $J=20\, T_{c0}$, $\xi_{s}=\xi_{f}$, $d_{s}= 2 \xi_{s}$, $\gamma=0.15$ and $\gamma_{b}=0$.} \label{Tcdf}
\end{center}
\end{figure}

Henceforth, we discuss only the physically relevant solution $p=0$. The superconducting critical temperature $T_{c}$ clearly increases with $Q$ as shown in the Fig.~\ref{FigTcp} (see also the Fig.~2 and Fig.~3 of Ref.~\onlinecite{Cha2004}). This enhancement is observed for a large range of parameters we have tested,
and thus is likely to be general.
The dependence of $T_{c}$ as a function of $d_{f}$ is shown in Fig.~\ref{Tcdf} for two different values of the dimensionless parameter $Q \xi_{s}$ (here we took $J=20\, T_{c0}$, $\xi_{s}=\xi_{f}$, $d_{s}= 2 \xi_{s}$, $\gamma=0.15$ and $\gamma_{b}=0$).
An important characteristic feature revealed by this figure is the evolution of the nonmonotonic dependence of $T_{c}(d_{f})$ in favor of a monotonic one in the presence of a moment inhomogeneity in the ferromagnet.

As an exacerbation of this nonmonotonic behavior of $T_{c}(d_{f})$, it is well-known \cite{Fom2002} that  within a particular choice of parameters the superconducting critical temperature may even jump to zero  for a finite range of thicknesses $d_{f}$ with a reentrance of superconductivity for higher values of $d_{f}$.
We investigated the influence of the moment inhomogeneity within this particular range of $d_{f}$ where $T_{c}=0$.
 As shown in the Fig.~2 of Ref.~\onlinecite{Cha2004}, we found 
the restoration of the superconductivity above a threshold value for the spiral wavevector $Q$. 
It is worth mentioning that a similar reentrance of superconductivity caused by a moment inhomogeneity is known in some rare-earth intermetallic compounds where a magnetic order competes with a superconducting order (see Ref.~\onlinecite{Bul1985}). In these compounds superconductivity is accompanied by a transition from the homogeneous ferromagnetic state to the cryptoferromagnetic state \cite{Bul1985}. However, it is important to note that this similar reentrance behavior in bulk systems and in S/F hybrid structures does not result from a similar physical mechanism.
In S/F  structures this behavior  results from a non-local influence of the magnetic order on the superconducting order (since both orders are assumed to be spatially separate) based on the Andreev reflection process.
According to the Sect. II, this process is characterized by the
coexistence of singlet and triplet pair correlations near the S/F interface.

\begin{figure}[b]
\begin{center}
\includegraphics[height=6.5cm]{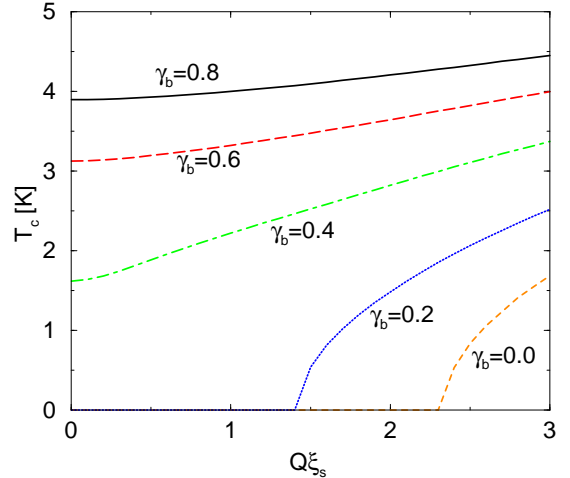}
\caption{$T_{c}$ versus the spiral wavevector $Q$ for different values of the interface parameter $\gamma_{b}$. Here $d_{f}=5$ nm and $d_{s}=7$ nm. The remaining parameters are the same as in Fig.~2 and 3 of Ref.~\onlinecite{Cha2004}.} \label{Figgb}
\end{center}
\end{figure}

In the Fig.~\ref{Figgb} we have represented the dependences 
$T_{c}(Q)$ for different values of 
the
 parameter $\gamma_{b}$ (the other parameters are the same as in Fig.~2 and Fig.~3 of Ref.~\onlinecite{Cha2004}).
The threshold value for $Q$ corresponding to the switch from the normal state to the superconducting state depends as expected on the quality of the S/F interface characterized here by $\gamma_{b}$ (in the case of a perfect transparency at the interface  $\gamma_{b}=0$, while $\gamma_{b} \gg 1$ for a low-barrier transparency). From the Fig.~\ref{Figgb}, one sees 
 that a relatively high transparency of the S/F interface is needed in order to have a strong dependence of $T_{c}$ on the wavevector $Q$.

\begin{figure}[t]
\begin{center}
\includegraphics[height=6cm]{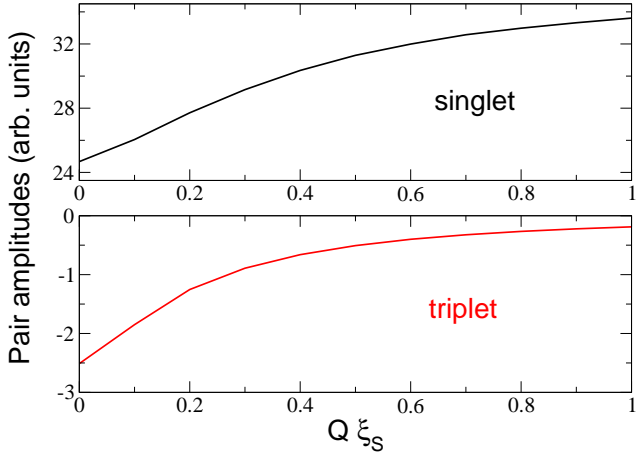}
\caption{Singlet and triplet pair amplitudes at the S/F interface versus the spiral wavevector $Q$. An increase of $Q$ is accompanied by an enhancement of the singlet amplitude together with a reduction of the (negative) triplet amplitude.
Here, we took the same parameters as in Fig.~\ref{Tcdf} with $d_{f}=2 \xi_{s}$.} \label{PairQ}
\end{center}
\end{figure}

So far we have only discussed the evolution of $T_{c}$ with the exchange field inhomogeneity. In the following we show how
the singlet and triplet pair amplitudes depend on the spiral wavevector $Q$.
For this purpose, we study the quantities
\begin{eqnarray}
{\cal F}_{s}(z)&=&T \sum_{n \geq 0}f_{s}(\omega_{n},z) \nonumber \\
{\cal F}_{tZ}(z)&=&T \sum_{n \geq 0} \mathrm{Im} \, f_{tZ}(\omega_{n},z) \nonumber.
\end{eqnarray}
In the Fig.~\ref{PairQ}, one sees that the enhancement of $T_{c}$ with $Q$ can be related with the enhancement of the singlet amplitude ${\cal F}_{s}(0)$ at the interface. 
We considered the parameter $\gamma_{b}=0$, which means that the singlet and triplet pair amplitudes are continuous at the interface. 
The triplet amplitude ${\cal F}_{tZ}(0)$ penetrating the singlet superconductor (which is negative in the plot) decreases to zero  when $Q$ increases. Thus, it seems that a reduction of the triplet pair correlations is in correspondence with an enhancement of the superconducting critical temperature $T_{c}$.

\begin{figure}[t]
\begin{center}
\includegraphics[height=6cm]{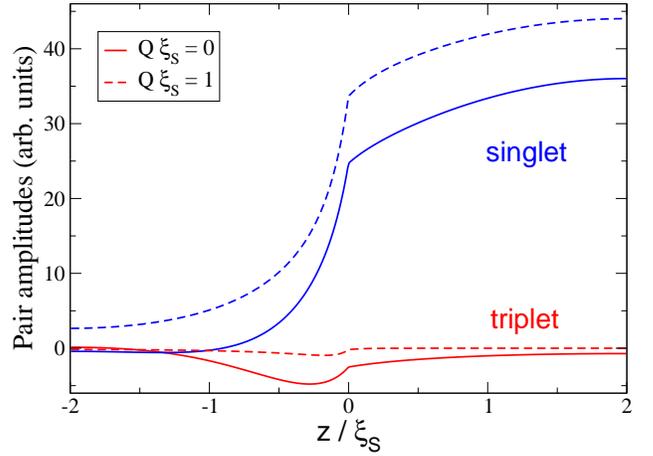}
\caption{Spatial dependence of singlet and triplet amplitudes for $Q \xi_{s}=0$ and $Q \xi_{s}=1$. Here we used the same parameters as in Fig.~\ref{Tcdf} with $d_{f}=2 \xi_{s}$.} \label{DepSpat}
\end{center}
\end{figure}

The most important effect of the inhomogeneity concerns the features of the spatial dependence of the pair amplitudes. It is known that a fixed exchange field may give rise to a change of sign of the
pair amplitude in the ferromagnet at some distances of the S/F interface (this occurs under specific conditions, e.g. a sufficiently large $d_{f}$). Actually, both singlet and triplet amplitudes may have a non-monotonic spatial dependence and change sign.
This feature is illustrated in Fig.~\ref{DepSpat}: the curves for the singlet and triplet correlations change both sign in the ferromagnet at $Q=0$. One can even note that they intersect near the position $z= - 1.3 \, \xi_{s}$. 
As shown in Fig.~\ref{DepSpat}, the shape of the spatial dependences for the singlet and triplet amplitudes is modified in the presence of the inhomogeneity ($Q \neq 0$). The crossing of the singlet and triplet curves tends to be suppressed with $Q$. Above a threshold value for $Q$, one finds that the singlet and triplet amplitudes do not change sign.

Finally, it is possible to capture the numerical results concerning the reentrance of $T_{c}$ or the suppression of the nonmonotonic spatial dependence of the pair amplitudes
with the exchange field inhomogeneity  at a qualitative level.
In fact, these features  stem from the imaginary character of the eigenvalues $k_{f}^{2}$ which are for $Q=0$
\begin{equation} 
k_{f\pm}^{2}=2\left( \omega_{n}\pm i J\right)/D_{f}
.\end{equation}
In Section III.B. we obtained  that in the presence of the inhomogeneity, $J$ has to be replaced by an effective exchange field $\tilde{J}_{\pm}(Q)$
\begin{equation}
J \to \tilde{J}_{\pm}(Q)=J \left( \sqrt{1-\eta^{2}} \mp i \eta \right)
\end{equation}
where we remind that $\eta=D_{f}Q^{2}/4J$. It is worth noting that now the imaginary part of $k_{f}^{2}$ for $\eta \leq 1$  is proportional to $\sqrt{1-\eta^{2}}$. Obviously, it is reduced in the presence of the inhomogeneity of the exchange field ($\eta \neq 0$) compared to the value obtained for a fixed exchange field ($\eta=0$). Furthermore, for $\eta \geq 1$, the eigenvalues $k_{f\pm}^{2}$ become only real.
 This indicates that the inhomogeneity of the exchange field
 is detrimental to all the interesting features that characterize the S/F proximity effect such as the non-monotonic dependence $T_{c}(d_{f})$, the oscillatory behavior of the pair amplitudes or of the local density of states in the ferromagnet.

\section{Conclusion}

We have discussed the system of spin-dependent quasiclassical equations 
that describes diffusive S/F hybrid structures.
We have pointed out that it is possible to predict many physical features of the S/F proximity effect 
directly from 
these equations, 
independent of
the specific geometry for the hybrid structure or the
specific model for the spatial evolution of the exchange field in the ferromagnet.
Singlet and triplet pair correlations are demonstrated to always coexist near the S/F interface.
The triplet vector has the tendency to be aligned with the exchange field, resulting predominantly into
 triplet pairs with zero-spin projection.
We have shown that the triplet correlations with nonzero spin projection on the local exchange field,
which may have a long-range penetration in the ferromagnet, arise when perfect alignment between the triplet 
vector with the magnetic moment is prevented. 
The production of triplet components is accompanied by a spin-splitting in the local density of 
states and by the induction of a spin magnetization near the S/F interface.

We have studied quantitatively the dependences of the singlet and triplet pair amplitudes
and of the superconducting critical temperature on the spiral wavevector within a model of a 
rotating magnetization in the ferromagnet.
The typical non monotonic behaviors of these quantities present for bilayers with homogeneously magnetized ferromagnets
are suppressed with increasing degree of inhomogeneity.
This demonstrates the influence of domain walls in the ferromagnet on 
measurable properties, implying that it is 
necessary to characterize the domain structure in order to compare quantitatively experiment with theory.
The dependence of the superconducting properties on the inhomogeneity of the 
exchange field
is expected to be a general feature of the proximity effect in mesoscopic hybrid structures 
composed of superconductors and ferromagnets, because it is a signature of the triplet correlations 
with zero-spin projection induced near the S/F interface.

\section{Acknowledgements}

We thank T. L\"{o}fwander for stimulating discussions.
This work was supported by the Deutsche Forschungsgemeinschaft within the Center for Functional Nanostructures.

\end{document}